\documentclass[useAMS,usenatbib]{mn2e}
\usepackage{epsfig,natbib}
\usepackage{amssymb}
\usepackage{graphicx}
\usepackage{placeins}
\usepackage{times}
\usepackage{epsfig}
\usepackage{amsmath}

\newcommand\ion[2]{#1{\sc #2}}

\title[A transition in the mode of galaxy growth at $z=2.6$]
{Mass-metallicity relation from z=5 to the present:
Evidence for a transition in the mode of galaxy growth at z=2.6 due
to the end of sustained primordial gas infall}
\author[P. M{\o}ller et al.]{
P. M{\o}ller,$^{1}$\thanks{E-mail: pmoller@eso.org}
J. P. U. Fynbo,$^{2}$
C. Ledoux,$^{3}$
K. K. Nilsson,$^{1}$ \\
$^{1}$European Southern Observatory, Karl-Schwarzschildstrasse 2, 85748
Garching bei M\"unchen, Germany\\
$^{2}$Dark Cosmology Centre, Niels Bohr Institute, Copenhagen
University, Juliane Maries Vej 30, 2100 Copenhagen O, Denmark\\
$^{3}$European Southern Observatory, Alonso de C\'ordova 3107, Vitacura,
Casilla 19001, Santiago 19, Chile
}

\begin{document}


\pagerange{1 -- 1} \pubyear{2011}

\maketitle


\begin{abstract}
We analyze the redshift evolution of the mass-metallicity relation
in a sample of 110 Damped Ly$\alpha$ absorbers spanning the redshift
range $z=0.11-5.06$ and find that the zero-point of the correlation
changes significantly with redshift. The evolution is such that the
zero-point is constant at the early phases of galaxy growth (i.e. no
evolution) but then
features a sharp break at $z=2.6\pm 0.2$ with a rapid incline towards
lower redshifts such that damped absorbers of identical masses are more
metal rich at later times than earlier. The slope of this mass
metallicity correlation evolution is $0.35 \pm 0.07$ dex per unit
redshift.

We compare this result to similar studies of the redshift evolution
of emission selected galaxy samples and find a remarkable agreement
with the slope of the evolution of galaxies of stellar mass
log$(M_{*}/M_\odot) \approx 8.5$. This allows us to form an
observational tie between damped absorbers and galaxies seen in
emission.

We use results from simulations to infer the virial mass of the
dark matter halo of a
typical DLA galaxy and find a ratio $(M_{vir}/M_{*}) \approx 30$.

We compare our results to those of several other studies that have
reported strong transition-like events at redshifts around
$z=2.5-2.6$ and argue that all those observations can be understood as
the consequence of a transition from a situation where galaxies were
fed more unprocessed infalling gas than they could easily consume
to one where they suddenly become infall starved and turn to
mainly processing, or re-processing, of previously acquired gas.

\end{abstract}


\section{Introduction}

The conversion of gas into stars and the ensuing enrichment of the
primordial gas with heavier elements is a fundamental process of
cosmic evolution, its monotonic nature making it akin to an arrow
of time for that evolution. Heavy
elements are at high redshifts most easily detected in absorption and
Damped Lyman $\alpha$ absorbers (DLAs, \citet{Wolfe86})
allow us to trace the metallicity of the cold
gas in galaxies back to redshifts of $z\gtrsim5$ 
\citep[e.g.,][]{Lu96,Prochaska03,Rafelski12}. Initially it was
thought that DLA studies directly would provide us with the history of
metal enrichment and that models based on DLA samples would be able to
fully describe this evolution \citep[e.g.,][]{Pei95,Cen99}.
However, the data displayed a surprisingly weak dependence of the DLA
metallicities on redshift \citep[for a review see][]{Pettini06}.  
Also, the population of Lyman-break galaxies (LBGs) were
discovered and some of these were found to have significantly higher 
metallicities than found in the DLA systems at similar redshifts
\citep[e.g.,][]{Pettini02}. The relation between DLAs and LBGs was
subsequently the subject of several studies
\citep{Fynbo99,Bunker99,Warren01,Moller02,Weatherley05,
Rauch08} and it was found
that the difference between the two sets of galaxies can
be understood from their very different selection functions
(gas cross-section vs. UV luminosity selection). Also, it is becoming
increasingly clear that correlations between metallicity and other
fundamental galaxy parameters like luminosity,
stellar mass, and star-formation rate are central to understanding the
galaxy population both locally and at high redshifts 
\citep[e.g.,][]{Tremonti04,Savaglio05,Calura09,Mannucci10}.
There is mounting evidence that DLA galaxies at
$z>2$ follow similar relations
\citep{Moller04,Ledoux06,Fynbo08,Pontzen08,Prochaska08,Krogager12}.

The Lyman break technique and searches for DLAs in quasar spectra
represent two widely different methods for finding galaxies at high
redshifts and as argued above they are now understood to select
different sub-samples of the high redshift galaxy population. Yet,
there is an overlap between the two which can be exploited to gain
a more detailed understanding of the galaxies in the overlapping
region.

Searches for high redshift Ly$\alpha$ emitters is a third and
independent method which again selects an independent well defined
sub-sample of the underlying sample of all high redshift galaxies.
\citet{Nilsson09,Nilsson11} found a remarkable property of Ly$\alpha$
emission selected galaxies namely an apparently sudden transition in
their dust properties between redshifts 2.3 and 3
\citep[confirmed by][]{Ciardullo12}. In an unrelated study
of highly reddened quasars \citet{Fynbo12} reported a
sudden drop in the detection rate of dusty quasar host galaxies at
redshifts above 2.6 in good agreement with the transition of dust
in Lyman $\alpha$ emitters at $z\simeq 2.5$.

In the same redshift range a transition in the mode of star formation
is seen to take place in galaxy formation models \citep{Oser10} and
\citet{Mannucci10} reported that here a
tight 3-way relation between mass, star formation
and metallicity appears for the first time.
This relation is thought to be caused by the interplay between infall
of pristine gas and the outflow of enriched gas. The fact that it
breaks down at $z\gtrsim3$ suggests that the balance between inflow and
outflow may be changing above and below this redshift, possibly caused
by differences in the initial mass function \citep{Bailin10} or
the thermodynamics of the intergalactic medium \citep{Dixon09}.

\citet{Ledoux06} showed that there is a strong mass metallicity
relation for DLAs out to redshifts beyond $z=4$ but they also reported
tentative evidence that this relation was changing with redshift.
The aim of this paper is to further investigate and quantify if there
is indeed a redshift evolution of the DLA mass metallicity relation,
how this relates to the redshift evolution of the mass metallicity
relation of galaxies \citep{Maiolino08}
and in particular how this is related to the phase
transition in the physics of galaxies and star formation at a redshift
around 2.5-2.6 as described above.

\section{Redshift evolution of the metallicity mass relation}
\label{sec2}
\subsection{Sample definition}
\label{sample}
Our basis is the sample of 70 absorbers from \citet{Ledoux06} (their
Table~1). In order to increase the redshift coverage over that sample
we have searched the literature for DLAs at both very low and very
high redshifts with published spectroscopic observations of
sufficiently high resolution. We found a total of 14 low redshift
DLAs (in the redshift range 0.1 - 1.5, Table~1) and we include
them all here. At high redshifts we selected the 22 DLAs observed
with HIRES from \citet{Rafelski12} which provides a homogeneous
high redshift sample providing good coverage back to $z=5.06$
(Table~2). In the sample we also included four newly observed objects
from our X-shooter DLA survey \citep{Fynbo10} since the parameters
required were already extracted and published (Table~3).
Our total sample therefore includes 110 systems sampling the redshift
range $z=0.11$ to 5.06. This represents an increase in sample size
of 60\% but more importantly it allows us to study the redshift
evolution over a large fraction of the age of the universe.

\begin{table*}
\caption{Metallicities and velocity widths of low-ionization lines for $z<1.5$ DLAs (metallicity measurements are from various papers quoted in the table footnote)}
\begin{tabular}{lccccccc}
\hline
\hline
Quasar & $z_{\rm abs}$ & $\log N($H\,{\sc i}$)$ & [X/H] &   & $\Delta V$    & Selected              & Refs. $^2$\\
       &               &                        &       & X & (km s$^{-1}$) & transition lines $^1$ &           \\
\hline
B\,0105$-$008  & 1.3710 &  $21.70 \pm 0.15$  &  $-1.40 \pm 0.16$ & Zn
&   31  & \ion{Cr}{\sc ii}$\lambda$2056 & a\\
J\,0256$+$0110 & 0.725  &  $20.70 \pm 0.20$  &  $-0.11 \pm 0.20$ & Zn
&  299  & \ion{Mg}{\sc i}$\lambda$2852 & b\\
Q\,0302$-$223  & 1.0094 &  $20.36 \pm 0.11$  &  $-0.56 \pm 0.12$ & Zn
&   61  & \ion{Si}{\sc ii}$\lambda$1808 & c\\
Q\,0449$-$1645 & 1.0072 &  $20.98 \pm 0.06$  &  $-0.96 \pm 0.08$ & Zn  &  204  & \ion{Al}{3}$\lambda$1862 & d\\
Q\,0454$+$039  & 0.8597 &  $20.69 \pm 0.06$  &  $-1.01 \pm 0.12$ & Zn
&  111  & \ion{Fe}{\sc ii}$\lambda$2260 & c\\
Q\,0933$+$733  & 1.479  &  $21.62 \pm 0.10$  &  $-1.58 \pm 0.10$ & Zn
&   30  & \ion{Fe}{\sc ii}$\lambda$2249 & e\\
Q\,0948$+$433  & 1.233  &  $21.62 \pm 0.05$  &  $-1.14 \pm 0.06$ & Zn
&   65  & \ion{Fe}{\sc ii}$\lambda$2260 & e\\
J\,1009$+$0713 & 0.1140 &  $20.68 \pm 0.10$  &  $-0.62 \pm 0.16$ & S
&   56  & \ion{Ti}{\sc ii}$\lambda$3384 & f\\
J\,1107$+$0048 & 0.740  &  $21.00 \pm 0.04$  &  $-0.54 \pm 0.16$ & Zn
&  212  & \ion{Mg}{\sc i}$\lambda$2852 & b\\
Q\,1354$+$258  & 1.4200 &  $21.54 \pm 0.06$  &  $-1.61 \pm 0.16$ & Zn
&   25  & \ion{Fe}{\sc ii}$\lambda$2260 & g\\
J\,1431$+$3952 & 0.6018 &  $21.2  \pm 0.1 $  &  $-0.80 \pm 0.21$ & Zn
&   58  & \ion{Mn}{\sc ii}$\lambda$2576 & a\\
J\,1623$+$0718 & 1.3357 &  $21.35 \pm 0.10$  &  $-1.07 \pm 0.13$ & Zn
&   59  & \ion{Fe}{\sc ii}$\lambda$2260 & a\\
J\,2328$+$0022 & 0.652  &  $20.32 \pm 0.06$  &  $-0.49 \pm 0.16$ & Zn
&   97  & \ion{Mg}{\sc i}$\lambda$2852 & b\\
B\,2355$-$106  & 1.1723 &  $21.0  \pm 0.1 $  &  $-0.87 \pm 0.20$ & Zn
&  131  & \ion{Mg}{\sc i}$\lambda$2852 & a\\
\hline
\end{tabular}
\flushleft
$^1$ Transition lines used to determine the velocity widths of low-ionization line profiles.\\
$^2$ References: [a] \citet{Ellison12};
[b] \citet{Peroux06};
[c] \citet{Pettini00};
[d] \citet{Peroux08};
[e] \citet{Rao05};
[f] \citet{Meiring11};
[g] \citet{Pettini99}.
\end{table*}

\begin{table*}
\caption{Metallicities and velocity widths of low-ionization lines for
$z\sim4$ DLAs (metallicity measurements are from \citet{Rafelski12} based on element X
with an upward correction of 0.3 dex for Fe)}
\begin{tabular}{lcccccc}
\hline
\hline
Quasar & $z_{\rm abs}$ & $\log N($H\,{\sc i}$)$ & [X/H] &   & $\Delta V$    & Selected             \\
       &               &                        &       & X & (km s$^{-1}$) & transition lines $^1$\\
\hline
J\,0040$-$0915 & 4.7394 &  $ 20.30 \pm 0.15 $  &  $ -1.40 \pm 0.17 $ &
Fe$+0.3$ &   67     & \ion{Fe}{\sc ii}$\lambda$1608 \\
J\,0747$+$4434 & 4.0196 &  $ 20.95 \pm 0.15 $  &  $ -2.28 \pm 0.20 $ &
Fe$+0.3$ &   69     & \ion{Fe}{\sc ii}$\lambda$1608 \\
J\,0817$+$1351 & 4.2584 &  $ 21.30 \pm 0.15 $  &  $ -1.15 \pm 0.15 $ &
S        &   89     & \ion{S}{\sc ii}$\lambda$1250  \\
J\,0825$+$3544 & 3.2073 &  $ 20.30 \pm 0.10 $  &  $ -1.68 \pm 0.16 $ &
Fe$+0.3$ &   30     & \ion{Fe}{\sc ii}$\lambda$1608 \\
J\,0825$+$3544 & 3.6567 &  $ 21.25 \pm 0.10 $  &  $ -1.83 \pm 0.13 $ &
Si       &   89/138 $^2$ & \ion{Si}{\sc ii}$\lambda$1808/\ion{Si}{\sc
ii}$\lambda$1304 \\
J\,1051$+$3107 & 4.1392 &  $ 20.70 \pm 0.20 $  &  $ -1.99 \pm 0.21 $ &
S        &   87     & \ion{Si}{\sc ii}$\lambda$1304 \\
J\,1051$+$3545 & 4.3498 &  $ 20.45 \pm 0.10 $  &  $ -1.88 \pm 0.10 $ &
Si       &   29     & \ion{Si}{\sc ii}$\lambda$1808 \\
J\,1051$+$3545 & 4.8206 &  $ 20.35 \pm 0.10 $  &  $ -2.28 \pm 0.10 $ &
Si       &   39     & \ion{Si}{\sc ii}$\lambda$1526 \\
J\,1100$+$1122 & 4.3947 &  $ 21.74 \pm 0.10 $  &  $ -1.68 \pm 0.18 $ &
Fe$+0.3$ &   137    & \ion{Ni}{\sc ii}$\lambda$1709 \\
J\,1200$+$4015 & 3.2200 &  $ 20.85 \pm 0.10 $  &  $ -0.64 \pm 0.10 $ &
S        &   145    & \ion{Ni}{\sc ii}$\lambda$1370 \\
J\,1200$+$4618 & 4.4765 &  $ 20.50 \pm 0.15 $  &  $ -1.38 \pm 0.16 $ &
Fe$+0.3$ &   88/113 $^2$ & \ion{Fe}{\sc ii}$\lambda$1611/\ion{Si}{\sc
ii}$\lambda$1304 \\
J\,1201$+$2117 & 3.7975 &  $ 21.35 \pm 0.15 $  &  $ -0.75 \pm 0.15 $ &
Si       &   490    & \ion{Si}{\sc ii}$\lambda$1808 \\
J\,1201$+$2117 & 4.1578 &  $ 20.60 \pm 0.15 $  &  $ -2.38 \pm 0.15 $ &
Si       &   25     & \ion{Si}{\sc ii}$\lambda$1526 \\
J\,1202$+$3235 & 4.7955 &  $ 21.10 \pm 0.15 $  &  $ -2.36 \pm 0.16 $ &
Fe$+0.3$ &   18     & \ion{C}{\sc ii}$^\star\lambda$1335 \\
J\,1202$+$3235 & 5.0647 &  $ 20.30 \pm 0.15 $  &  $ -2.66 \pm 0.16 $ &
Si       &   30     & \ion{O}{\sc i}$\lambda$1302  \\
J\,1304$+$1202 & 2.9131 &  $ 20.55 \pm 0.15 $  &  $ -1.65 \pm 0.16 $ &
S        &   64     & \ion{Fe}{\sc ii}$\lambda$1608 \\
J\,1304$+$1202 & 2.9289 &  $ 20.30 \pm 0.15 $  &  $ -1.54 \pm 0.16 $ &
S        &   22     & \ion{Fe}{\sc ii}$\lambda$1608 \\
J\,1353$+$5328 & 2.8349 &  $ 20.80 \pm 0.10 $  &  $ -1.38 \pm 0.10 $ &
S        &   43     & \ion{S}{\sc ii}$\lambda$1253  \\
J\,1438$+$4314 & 4.3990 &  $ 20.89 \pm 0.15 $  &  $ -1.31 \pm 0.15 $ &
S        &   91     & \ion{Fe}{\sc ii}$\lambda$1608 \\
J\,1541$+$3153 & 2.4435 &  $ 20.95 \pm 0.10 $  &  $ -1.49 \pm 0.11 $ &
Si       &   47     & \ion{Si}{\sc ii}$\lambda$1808 \\
J\,1607$+$1604 & 4.4741 &  $ 20.30 \pm 0.15 $  &  $ -1.71 \pm 0.15 $ &
Si       &   37     & \ion{Si}{\sc ii}$\lambda$1304 \\
J\,1654$+$2227 & 4.0022 &  $ 20.60 \pm 0.15 $  &  $ -1.66 \pm 0.16 $ &
Fe$+0.3$ &   22     & \ion{Fe}{\sc ii}$\lambda$1608 \\
\hline
\end{tabular}
\flushleft
$^1$ Transition lines used to determine the velocity widths of
low-ionization line profiles.\\
$^2$ For this system no line fulfilled the optimal criteria fully so we
measured both a very weak line and a strong (slightly saturated) line.
We list both lines and both measurements but we use the mean of the two.
\end{table*}

\begin{table*}
\caption{DLA sample: average metallicities and velocity widths of low-ionisation line profiles $^\ast$}
\begin{tabular}{llcccccccc}
\hline
\hline
Quasar & Other name & $z_{\rm em}$ & $z_{\rm abs}$ & $\log N($H\,{\sc i}$)$ & [X/H] &    & $\Delta V$    & Selected & Refs. $^2$\\
       &            &              &                    &                        &            & X  & (km s$^{-1}$) & transition lines $^1$\\
\hline
Q\,0151$+$048 &                               & 1.93 & 1.934 &
$20.34\pm 0.02$ & $-1.93\pm 0.04$ & Si &  51     & Si\,{\sc ii}$\lambda$1304 & a, b \\
Q\,0918$+$1636&{\small SDSS J 091826.16$+$163609.0}&3.07&2.583&$20.96\pm 0.05$ & $-0.12\pm 0.05$ & Zn & 295     & Si\,{\sc ii}$\lambda$1808 & c \\
Q\,0918$+$1636&{\small SDSS J 091826.16$+$163609.0}&3.07&2.412&$21.26\pm 0.06$ & $-0.55\pm 0.16$ & Zn & 352     & Si\,{\sc ii}$\lambda$1808 & d \\
Q\,2222$-$0946&  & 2.93 & 2.345 & $20.65\pm 0.05$ & $-0.46\pm 0.07$ & Zn & 185     & Si\,{\sc ii}$\lambda$1808 & e \\
\hline
\end{tabular}
\flushleft
$^1$ Transition lines used to determine the velocity widths of low-ionization line profiles.\\
$^2$ References: [a] \citet{Zafar11};
[b] \citet{Ellison12};
[c] \citet{Fynbo11};
[d] \citet{ThorsenThesis:2011};
[e] \citet{Fynbo10}.
\end{table*}

\subsection{Method}
\label{method}
There is a relation between the velocity width of a DLA system and the
gravitational mass of the host \citep{Haehnelt98,Pontzen08} which makes
it possible to use the velocity width, $\Delta V$, as a proxy for the
mass. In this paper we shall use the $\Delta V$ as defined in
\citet{Ledoux06} and for consistency we carefully choose suitable
low ionization
absorption lines as described in that work. To convert the published
high resolution line profile data to electronic data we used the
ADS Dexter data extraction applet which is a tool to extract data
from figures from the ADS article service \citep{Demleitner2001}.
The selected lines and the resulting $\Delta V$ are listed in Tables
1-2.

\citet{Ledoux06} found the 
slope of the metallicity-$\Delta V$ relation to be 1.46
and no evidence was found for an evolution of this slope with
redshift. We shall therefore at first conservatively adopt this
slope. This forms our conservative assumption, but
in order to certify that our conclusions do not
depend on this assumption we shall also repeat the entire
analysis with less conservative assumptions below.

The sample is plotted in the left panel of Fig.~1 where we also plot
the best fit linear relation for a slope of 1.46. The errors on
[M/H] are typically in the range $0.05-0.20$ and it is seen
\begin{figure}
\includegraphics[width=0.48\textwidth]{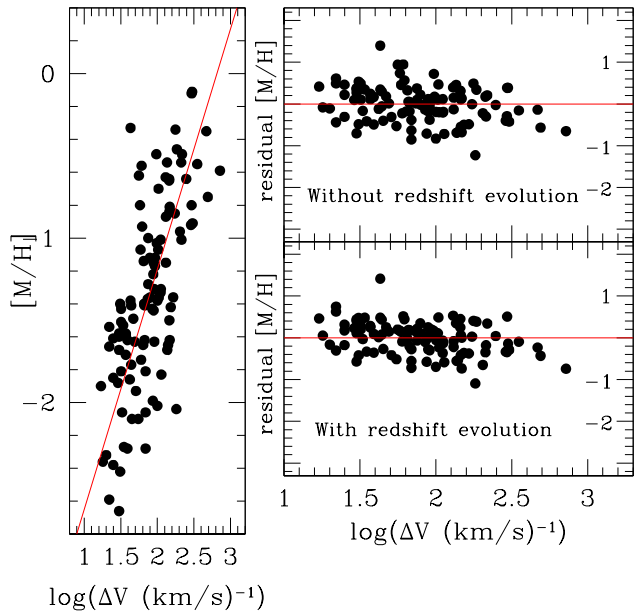}
\caption{{\it Left frame:} Metallicity versus velocity
width of each system in our sample.
The best fit linear relation using a slope of 1.46
(see text) is drawn as a red line. {\it Right frames top to bottom:}
Residuals after subtraction of the fit with no redshift evolution
(eq(2)) and after subtraction of the fit including redshift evolution
(eq(2 and 5)).}
\label{fig:Fig1}
\end{figure}
that the errors on the measurements are far too small to
account for the scatter. As a first step we shall now determine
the internal scatter of the distribution. To do this we assume
that the measurement errors on the metallicity ($\sigma_{\rm met}$)
as well as the natural internal scatter of the relation
($\sigma_{\rm nat}$) both follow normal distributions and that the two
are statistically independent. This allows us to define a total
predicted $\sigma$
\begin{equation}
\sigma_{\rm tot}(i) = (\sigma_{\rm met}^2(i) + \sigma_{\rm nat}^2)^{1/2}
\label{eq:1}
\end{equation}
for each observation, "$i$". For a given slope ($\alpha_{\rm 0}$)
and zero point ($zp$) of the metallicity-mass relation
\begin{equation}
[M/H] = \alpha_{\rm 0} ~ {\rm log} (\Delta V) + zp      
\label{eq:2}
\end{equation}
we can now assign a goodness of fit as the sum of
the squares of the deviations of the fit to the data
\begin{equation}
C_{\rm dof}^2 = \sum_{i=1}^{110} (([M/H](i) - \alpha_{\rm 0} ~ {\rm log}
(\Delta V(i)) - zp)/\sigma_{\rm tot}(i))^2)/dof 
\label{eq:3}
\end{equation}
where the degrees of freedom (dof) in this case is 108. ~
$C_{\rm dof}^2$ is equivalent to the $\chi ^2$ per degree of freedom
of the distribution
if $\sigma_{\rm nat}$ is correctly chosen. We can therefore
use eq(3) to select all pairs of ($zp$, $\sigma_{\rm nat}$) for which
$C_{\rm dof}^2$ = 1, and we
then choose the pair for which $\sigma_{\rm nat}$ has its minimum value
as the globally best fit:
($\alpha_{\rm 0}$, $zp$, $\sigma_{\rm nat}$) = (1.46, -4.111, 0.4181).
In the top right
panel of Fig.~1 we plot the residuals after subtraction of this fit.

\subsection{Redshift evolution}
By splitting their sample into a high and a low redshift sample,
\citet{Ledoux06} already pointed out that while the slope of the
correlation did not appear to change, there was a shift of the
correlation. Here we seek a more general formulation of this redshift
evolution.

Inverting eq(2) we can interpret each absorber of the sample as an
independent measurement of the zero point
\begin{equation}
zp(i) = [M/H](i) - \alpha_{\rm 0} ~ {\rm log} (\Delta V(i))
\label{eq:4}
\end{equation}
with an uncertainty of $\sigma_{\rm tot}$(i). This allows us to check
if we see an evolution of $zp$ as a function of redshift. The
function describes directly the evolution (shift) of the
mass-metallicity relation at a given intercept (mass) and we have
chosen to normalize the zero point to a mass of 100 km/s, which is a
typical mass in our sample, and
to name it $[M/H]_{100~km/s}(z)$. This function represents
the mean metallicity of a galaxy with a mass of 100 km/s as a function
of redshift.

In Fig.~2 we show $[M/H]_{100~km/s}(z)$ plotted versus redshift,
\begin{figure}
\includegraphics[width=0.40\textwidth]{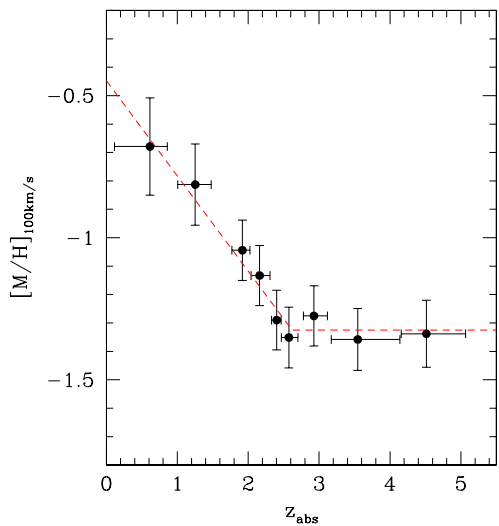}
\vskip -0.2cm
\caption{Redshift evolution of the zero point of the DLA
mass-metallicity relation. We have normalized the function to a
gravitational well mass of
100 km/s. The red line is a best fit to the data of the form given
in eq(5). There is no evidence for evolution before redshift 2.6,
and then a sharp break at $z=2.6$ with an onset of strong evolution.
The evolution is such that the metallicity of a galaxy of a given
mass grows by $0.35 \pm 0.07$ dex per unit redshift.
The fit was computed as a fit to all 110 data points, not to the
binned data.}
\label{fig:Fig2}
\end{figure}
i.e. we plot the evolution of the mean metallicity of the cold gas in
a galaxy of a mass of 100 km/s from redshift 5 to 0.
The horizontal bars in Fig.~2 show the $z$ range of each bin, while
the vertical bars show the propagated uncertainty of the
$[M/H]_{100~km/s}(z)$.

From Fig.~2 we see clear evidence for a steady and continuous incline
of $[M/H]_{100~km/s}(z)$ from redshift 2.6 to 0. An incline of $M(z)$
means that the
metallicity-mass relation changes such that for any given mass the
galaxy becomes more metal rich at later times. At earlier times than
redshift 2.6 there is no evidence for any evolution and it appears
that the mass-metallicity relation was constant from $z=5$ to $z=2.6$.

We will take this apparent ``two phase'' nature of the early universe
as our working hypothesis and proceed to fit the evolution with a
constant at high redshifts and the late incline with a linear
function. The transition between the two then happens
at the transition redshift $z_{\rm tran}$ and we get
\begin{equation}
zp(z) =
\begin{cases}
\beta_{\rm early} (const)   & \text{for }   z>z_{\rm tran},\\
\alpha_{\rm late}~ z + \beta_{\rm late} & \text{for } z\leq z_{\rm tran}
\end{cases}\label{eq:5}
\end{equation}
where the constant value at high redshift is defined by the requirement
that $zp(z)$ is continuous at $z_{\rm tran}$, i.e. $\beta_{\rm early}$
is not a free parameter.
\begin{table}
\caption{Fitted evolution parameters for $\alpha_{\rm 0}=1.46$.
The parameters are defined in eq(5).}
\begin{tabular}{lc}
\hline
\hline
$\alpha_{\rm late}$ & $-0.35 \pm 0.07$ \\
$\beta_{\rm late}$ & $-3.33 \pm 0.16$ \\
$z_{\rm tran}$ & $2.62 \pm 0.20$ \\
$\beta_{\rm early}$ & $-4.25 \pm 0.05$ \\
\hline
\end{tabular}
\end{table}

We do not fit the binned data in Fig.~2, rather we proceed using the
method described in Sect.~2.2 fitting simultaneously the parameters in
eq(2) and eq(5) for all 110 systems. In this way we find
$z_{\rm tran}$=2.618
and $\sigma_{\rm nat}$=0.3759 for $\alpha_{\rm 0}$=1.46 and 106 dof.
In Table~4 we provide all the parameters for the fit and in
Fig.~2 this best fit is shown as red dashed lines.
In Fig.~1 (lower panel to
the right) we plot the residuals ($res_{\rm [M/H]}$)
after subtraction of this fit. Comparing the upper and lower right
frame it is easy to notice the decreased scatter of the residuals
(from 0.4181 to 0.3759).

\subsection{Testing the importance of the assumed $\alpha_{\rm 0}$}

In the above analysis we started by fixing the slope of the mass
metallicity relation to $\alpha_{\rm 0}$=1.46 as found by
\citet{Ledoux06} who used a bisector fit. We agree that because of the
large scatter the bisector is the best way of determining the slope,
but we wish to test if another assumption would significantly
change our result regarding the redshift evolution. We therefore now
relax this constraint and repeat the analysis treating
$\alpha_{\rm 0}$ as a free parameter with an additional minimization
fit. We find that the best fit in the form of eq(2) is obtained with a
slope of $\alpha_{\rm 0} = 1.12$ as shown in Fig.~3. The resulting
fit to the redshift evolution is however almost unchanged. The
transition redshift is now found at $z_{\rm tran}$=2.622 (a change of
only 0.003) and the redshift evolution slope is 0.37 dex per unit
redshift (a change of 0.02). Both values are well within the
$1 \sigma$ errors of the original fit.
\begin{figure}
\includegraphics[width=0.48\textwidth]{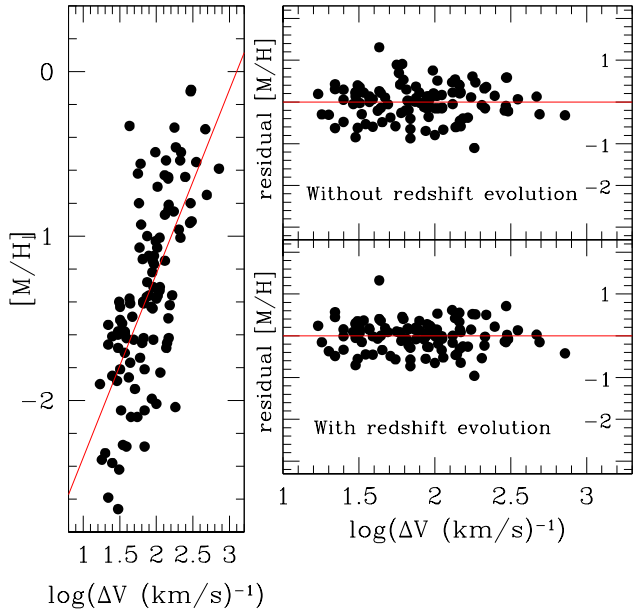}
\caption{Same as Fig.~1 except that we here fit for the slope
and find a best fit for $\alpha_0 = 1.12$.}
\label{fig:Fig3}
\end{figure}

We conclude from this that the specific assumption of the slope
of the underlying mass metallicity relation has almost no effect on
the results regarding the redshift evolution of the relation and
that the result displayed in Fig.~2 is very robust.
In hindsight we can now understand the riddle which for almost
2 decades was perplexing cosmologists:  ``Why do we not see clear
evidence for cosmic chemical evolution in high redshift DLAs?''
\citep[e.g.,][]{Pettini94}.

\subsection{Comparison with the redshift evolution of the galaxy
stellar mass-metallicity relation}

\citet{Maiolino08} provided galaxy mass-metallicity relations back to
a redshift of 3.5 as a function of galaxy stellar mass (their
eq.~2). We have computed a grid of those relations and overplot them
on our DLA measurements in Fig.~4. It can be seen that
\citet{Maiolino08} produces relations of metallicity as a function
of redshift of increasing slopes with decreasing stellar masses. 
In the mass range
log$(M_{*}/M_\odot )=8.5-9.0$, it is apparent that the slope matches
the evolution of DLAs below $z\approx 2.6$ almost perfectly. The fact
that the evolution of the zero point of the DLA velocity-metallicity
correlation with redshift matches the galaxy stellar mass-metallicity
evolution so well strongly supports the idea that we are witnessing
in the DLA data a mass-metallicity relation akin to galaxies as
initially suggested by \citet{Ledoux06} further.
\begin{figure}
\includegraphics[width=0.48\textwidth]{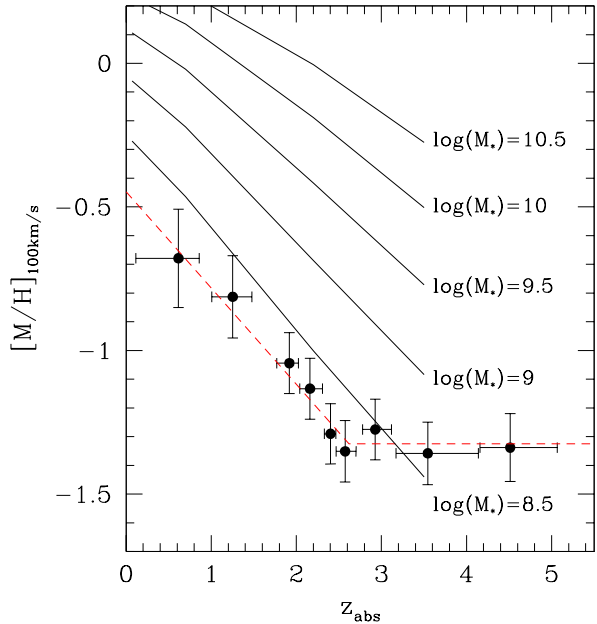}
\caption{Here we plot the redshift evolution of the galaxy
mass-metallicity relations from \citet{Maiolino08} for different
stellar masses (full curves) and the DLA data points and fit (dashed
line) from Fig.~2. There is a remarkable agreement between the slope
of the redshift evolution of the DLA galaxies and that of low-mass
galaxies from Maiolino's study.}
\label{fig:Fig4M}
\end{figure}

It is still a matter of debate how exactly emission-line based
metallicities relate to the metallicity of the absorbing gas. It is
usually thought that the DLA gas can have lower metallicities than
those determined from emission line fluxes \citep[e.g.][]{Peroux12}.
However, there are uncertainties in the recipes used to measure
oxygen abundances from emission lines \citep[e.g.,][]{Kudritzki12}
and in the choice of a
suitable element to derive absorption-line metallicities. Other works
have indeed found consistency between the two measurements 
\citep[e.g.,][]{Bowen05,Krogager13}. Here, we conservatively assume
that the DLA gas either has the same metallicity or a metallicity up
to 0.5 dex lower than that derived from emission lines.
Any effects related to $\alpha$-element overabundance will also
be covered by this assumption.  Given the
remarkable similarity of the slopes of the relations shown in
Fig.~\ref{fig:Fig4M}, we then estimate that the ``typical'' DLA in
our sample exhibiting $\Delta V=100$~km s$^{-1}$ has a stellar mass
log$(M_{*}/M_\odot )$ of the order of 8.5. This is in agreement with
the early result that due to the effect of gas cross-section selection
even the most massive DLA galaxies only correspond to the least
massive Lyman-Break galaxies \citep{Fynbo99}. As a matter of fact,
there is only a small range of overlapping metallicity between our
DLA sample and the galaxy sample in the study of \citet{Maiolino08}.

\begin{figure}
\includegraphics[width=0.48\textwidth]{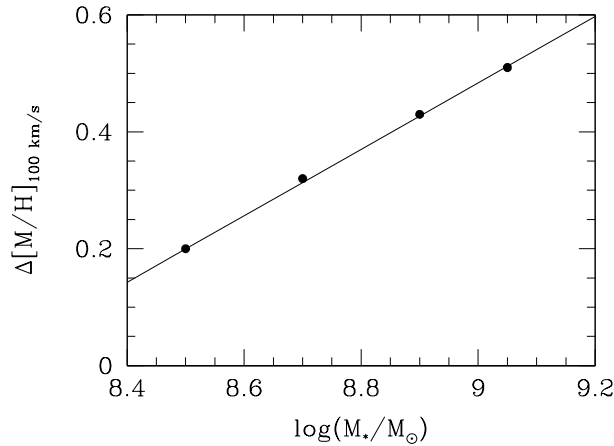}
\vskip -2.6cm
\caption{
Offsets ($\Delta$[M/H]$_{100 {\rm km/s}}$) required to fit
the data points in Fig.~4 to a grid of stellar masses using the
Maiolino models in the region of overlap. The offsets are well
fitted by a linear relation with slope $1.76^{-1}$.}
\label{fig:Fig5}
\end{figure}

In order to generalize the comparison, we then determined the best fit
to a range of models in the very lowest range of masses covered by
Maiolino's data, which constitutes the small overlap region with the
DLA galaxies. In practice we created a finer grid of the Maiolino
models than shown in Fig.~4, we then slid our data points in Fig.~4
up to match each model defined by a given stellar mass and
determined the offset ($\Delta$[M/H]$_{100 {\rm km/s}}$) required.
The offsets are shown in Fig.~5 as a function of stellar mass and it
is seen that in the mass range considered the offsets are well
fitted by a linear relation with slope $1.76^{-1}$.

The correction in metallicity when comparing absorption to emission
is not yet well known so we allow for this as a free parameter,
C$_{[M/H]}$, which must be added to the absorption-line metallicity
to make it consistent with the emission-line metallicity. Finally,
in the redshift range considered we found an evolution with redshift
such that $\Delta$[M/H]$=-0.35\Delta z$ for a given mass.
We can now write up the relation:
\begin{equation}
\log (M_{*}/M_\odot ) = 1.76 ([M/H] + C_{[M/H]} + 0.35 z + 5.04)
\label{eq:Maiolino}
\end{equation}
between DLA metallicity and stellar mass of the DLA host galaxy. Here
1.76 on the right side comes from the slope of the relation in Fig.~5,
$0.35 z$ from the redshift evolution ($\alpha_{\rm late}$ in eq(5))
and 5.04 is the empirically determined zero point of the relation.

A number of important caveats should be noted before using this
relation.
\begin{itemize}
\item
There is a large scatter in this relation (0.38 dex in $[M/H]$) so it
is only useful as a statistical tool on large samples. Application to
single systems will probably not be meaningful.
\item
The relation is found from, and only known to be correct in, the small
range of overlap between the two samples.
\item
The relation is formally valid back to $z=2.6$ and the fit is
excellent between $z=2.6$ and $z=0$. At higher redshifts, flux-limited
galaxy studies only sample the most massive galaxies, while the DLA
systems sample halos of roughly the same mass at all redshifts. There
is therefore a huge gap between the two samples at high redshifts and
we have not tried to extract a similar relation past the break at
$z=2.62$. The results provided in this paper suggest that the relation
remains constant at higher redshifts (i.e., $z$ should be replaced by
2.62 in Eq.~\ref{eq:Maiolino}), but this has not yet been verified by
high-redshift galaxy studies.
\end{itemize}

\subsection{Virial masses}

We now ask the question what the virial mass ($M_{\rm vir}$) of the
dark matter halo of a typical DLA galaxy is. Via identification of
DLAs in dark matter halos of their cosmological simulations,
\citet{Pontzen08} were able to confirm that the relation between
$\Delta V$ and metallicity reported by \citet{Ledoux06} is indeed a
mass-metallicity relation. A relation between $\Delta V$ and virial
mass of the DM halo is clearly seen in their simulations but the
relation has a significant scatter. They also tend to underestimate
the velocity widths of absorption-line profiles (their fig.~8). Taking
this into account, a typical DLA with width 100~km s$^{-1}$ would
correspond to a virial mass log$(M_{\rm vir}/M_\odot )$ of the order
of 10. This leads to a mass-to-light ratio of $\sim 30$ for a typical
DLA galaxy at redshifts below $z\approx 2.6$.
We note that significantly larger virial masses have been 
inferred for DLA host halos from clustering arguments
\citep{Font-Ribera12}.

\section{Discussion and conclusions}

\subsection{Transition redshift of the mass metallicity relation}

We confirm the reported mass metallicity relation for high redshift
DLAs and we confirm that there is indeed a redshift evolution of this
relation \citep{Moller04,Ledoux06,Pontzen08} but only in the later
phase of cosmic evolution. At high redshifts the mass metallicity
relation is constant while there is a break at $z = 2.6 \pm 0.2$
with a sudden onset of redshift evolution with a slope of
$0.35 \pm 0.07$ dex per unit redshift.

Previously similar sharp transitions in the properties of Ly$\alpha$
emitters identified as ULIRGs and QSOs
\citep{NilsMoll09,Bongiovanni10,NilsMoll11} and in the detection rate of
highly reddened quasar host galaxies \citep{Fynbo12} at redshifts
around $z=2.5-2.6$ have been reported.
The close coincidence of the redshifts of those transitions, which also
precisely coincides with the peak of the star formation rate density
\citep{Hopkins06} makes it plausible that they must all be related in
some way.
Ly$\alpha$ emitters only form a sub-set of the population of high
redshift galaxies. DLAs however trace the gas both in Ly$\alpha$
emitting galaxies (e.g. \citet{Moller93}) and in galaxies with no
Ly$\alpha$ emission (e.g. \citet{Fynbo11}; \citet{Bouche12})
and in fact most DLA
galaxies are non-detected in Ly$\alpha$ \citep{Moller04}. It is
therefore now evident that the transition seen for Ly$\alpha$ emitters
is not an artefact of the Ly$\alpha$ emission selection method.
A recent detailed study of the mass metallicity relation of galaxies
seen in emission \citet{Mannucci10} showed that the relation in
fact forms a plane with the star formation rate as the third
parameter. Their description forms a perfect fit in the redshift
range $z=0-2.5$, but breaks completely down at z=3 as seen in
their Figure 4, right panel.

\subsection{Slope of the redshift evolution}
The redshift evolution below $z=2.6$ is such that for a given virial
halo mass the gas has higher
metallicity at lower redshifts than at higher redshifts. This is in
agreement with results for galaxies reported by \citet{Savaglio05}
and \citet{Maiolino08}.
We find the slope of the evolution to be $0.35\pm 0.07$ dex per
unit redshift in agreement with the slope found in the low stellar
mass range for similar studies of galaxies.

\subsection{The stellar masses of DLAs}
Identifying the galaxies that harbour the DLAs at high redshifts has
been a holy grail for DLA studies ever since the first DLA was
discovered. Attempts to do this on a case by case basis have been
mostly unsuccessful \citep[e.g.,][]{Smith89} and still only a couple
of handful detections have been reported \citep{Krogager12}.
In this paper we were able to tie a direct link between
a flux selected galaxy sample and our DLA sample using the evolution
of the mass-metallicity relation to link them, and to provide an
expression for the stellar mass of DLAs as a function of their
metallicity and redshift.

\subsection{A plausible cause of the transition: Infall starvation}

The conclusions from a study of the redshift evolution of Ly$\alpha$
emitters \citep{Nilsson09} were that at redshifts above $\sim3$ the
objects were all star forming, with little or no dust, and with
young blue populations, while at redshifts lower than $\sim2.5$ a
significant fraction of red, dusty, mixed population objects, even
ULIRGs, were found. The simplest way to explain those observations
as well as the new results reported here is a scenario in which
primordial, or near-primordial, gas is constantly accreted onto the 
galaxies at high redshifts \citep{Barkana03,Weidinger05,Dekel09}.
If the amount of gas available for infall is larger than a galaxy can
accept and process then they will all accrete at maximum capacity. As
the gas is converted into stars and enriches the galaxy with metals,
this sets up an unchanging universal relation between the mass and the
metallicity of a galaxy. With respect to this relation one might
refer to this early phase as a steady state of galaxy evolution.

In order to change the universal mass-metallicity relation, some other
parameter has to change first and the only available parameter
is the infall rate. We therefore propose that the trigger to
start the evolution of the mass-metallicity relation is a drop in the
infall rate of primordial gas at $z_{\rm tran} = 2.6$ below the point
of balance for the steady state to be maintained.

This can be
accomplished in two ways. Either the supply of metal poor gas simply
drops far below that of metal enriched gas, or at the peak of the star
formation rate at $z\simeq 2.6$ \citep{Hopkins06} powerful galactic
winds create highly ionized halos which render it difficult for the
infall to continue at the same high level.
This effect is known as accretion quenching and was used by
\citet{Bouche10} who found that it was a necessary ingredient in their
model in order to reproduce the observed star formation rate-mass and
Tully-Fisher relations. The primary mode of galaxy stellar mass growth
then shifts from gas accretion to merging. There is some
theoretical evidence for this.
\citet{Oser10} reported that their simulations showed that
there is a transition between redshifts of 3 and 2 where the
galaxy growth changes from ``in situ'' formation of stars from
accreted cold gas towards accretion of stars which were formed in
satellites earlier. The new results we have presented here are
observational support for this model as well as for the accretion
efficiency controlled model by \citet{Bouche10}.
There are similar obvious links to the sudden change in Ly$\alpha$
emission line selected samples at $z_{\rm tran} = 2.52$ and to the
setup of the tight 3-way relation between mass, star formation
and metallicity (FMR) between $z=3.3$ and 2.5. \citep{Mannucci10}.
In the latter work it was found that infall rates must have been
$\simeq 10^3$ times higher at $z=3.3$ than locally.

\subsection{Future work and outlook}

The dominating source of errors is the internal scatter, not the
measurement errors. Using a larger sample or a sample with smaller
errors will therefore not improve the results much. The best way
to obtain a significant improvement is to identify and 
understand the source of the scatter.

N-body simulations of the galaxy growth via ``in situ'' formation
of stars versus accretion of satellites including gas metallicity
predictions would be an important tool to test if the proposed
model is correct.

In order to be able to test if the break at $z=2.6$ is also present in
galaxy emission selected samples those would need to reach much lower
stellar mass at redshifts between 4 and 5 to make it possible to tie
them together with DLA samples. This will likely not be
possible until E-ELT class telescopes become available.

\section*{Acknowledgments}

The Dark Cosmology Centre is funded by the DNRF. JPUF acknowledges
support form the ERC-StG grant EGGS-278202.

\def\aj{AJ}
\def\araa{ARA\&A}
\def\apj{ApJ}
\def\apjl{ApJ}
\def\apjs{ApJS}
\def\apss{Ap\&SS}
\def\aap{A\&A}
\def\aapr{A\&A~Rev.}
\def\aaps{A\&AS}
\def\mnras{MNRAS}
\def\nat{Nature}
\def\pasp{PASP}
\def\aplett{Astrophys.~Lett.}

\bibliographystyle{mn}
\bibliography{thebib}

\end{document}